%% file: main.tex
\documentclass[a4paper,11pt]{article}
\pdfoutput=1 

\usepackage{jcappub} 

\usepackage{float}
\usepackage{subcaption}
\usepackage[
    input-ignore={.},
    group-separator={.},
    input-decimal-markers={,}, 
    output-decimal-marker={,},
    per-mode=power]{siunitx}
\usepackage{amsmath}
\usepackage{amsfonts}
\usepackage{amssymb}
\usepackage[capitalize]{cleveref}

\newcommand{\charon}{\raisebox{0.82\depth}{$\chi$}\texttt{aro$\nu$}}
\newcommand{\nunu}{\nu\bar{\nu}}
\newcommand{\bb}{b\bar{b}}
\newcommand{\tautau}{\tau^{-}\tau^{+}}

\begin{document}
\title{Sensitivity Projections for Low-Mass Dark Matter Annihilation with the IceCube Upgrade}

\date{\today}

\input{authors_only}

\newpage
\abstract{
The IceCube Upgrade, an extension designed to enhance the IceCube Neutrino Observatory's detection of neutrinos with energies between \qty{1}{\GeV} and \qty{500}{\GeV}, will markedly improve IceCube's sensitivity to low-mass dark matter scenarios.
In this study, we present sensitivity projections for the IceCube Upgrade to neutrino fluxes arising from dark matter annihilation.
In particular, we consider dark matter with masses between \qty{3}{\GeV} to \qty{500}{\GeV} from both the core of the Sun and the Galactic Center.
These projections indicate that the IceCube Upgrade will enable stringent limits on dark matter in this parameter space, achieving leading sensitivities to some dark matter models with only three years of data taking.
}

\maketitle

\input{introduction}
\input{IceCube_Upgrade}
\input{event_selection}
\input{analysis_method}
\input{results}
\input{conclusion}

\bigskip
\noindent
{\it Note added} --- While this document was being prepared for publication, the IceCube Collaboration has concluded the deployment of the IceCube Upgrade.
The Upgrade operates with five deployed strings out of the originally planned seven strings, and approximately 90\% of modules on these five strings have been successfully commissioned.
To estimate the impact of this reduced detector configuration on the sensitivity projections presented in this paper, hits on the affected DOMs were removed from the nominal MC simulation.
This modified simulation was then reprocessed through the full event selection pipeline, including detector triggering, reconstruction, and selection cuts, and the analysis was rerun on the resulting dataset.
However, the machine learning models used for noise cleaning, reconstruction, and event classification were not retrained for this configuration, nor were the selection cuts re-optimized; as such, this estimate is expected to be conservative.
The impact on the projected sensitivities varies across DM masses and annihilation channels, with the largest effect at the lightest dark matter masses.
In all cases, the changes remain within the $1\sigma$ expected statistical fluctuations of the analysis.
Additionally, the detector systematics were not re-run with the new configuration, but are expected to remain at the same level as before, well under the $1\sigma$ expected from statistical fluctuations.

\input{acknowledgments}

\input{appendix}

\input{references}

\end{document}

%% file: authors_only.tex
\author[15]{R. Abbasi,}
\author[62]{M. Ackermann,}
\author[16]{J. Adams,}
\author[9]{J. A. Aguilar,}
\author[20]{M. Ahlers,}
\author[21]{J.M. Alameddine,}
\author[34]{S. Ali,}
\author[42]{N. M. Amin,}
\author[40]{K. Andeen,}
\author[12]{C. Arg{\"u}elles,}
% ...
\affiliation[9]{Universit{\'e} Libre de Bruxelles, Science Faculty CP230, B-1050 Brussels, Belgium}
% ...
\affiliation[12]{Department of Physics and Laboratory for Particle Physics and Cosmology, Harvard University, Cambridge, MA 02138, USA}
% ...
\affiliation[15]{Department of Physics, Loyola University Chicago, Chicago, IL 60660, USA}
\affiliation[16]{Dept. of Physics and Astronomy, University of Canterbury, Private Bag 4800, Christchurch, New Zealand}
% ...
\affiliation[20]{Niels Bohr Institute, University of Copenhagen, DK-2100 Copenhagen, Denmark}
\affiliation[21]{Dept. of Physics, TU Dortmund University, D-44221 Dortmund, Germany}
% ...
\affiliation[34]{Dept. of Physics and Astronomy, University of Kansas, Lawrence, KS 66045, USA}
% ...
\affiliation[40]{Department of Physics, Marquette University, Milwaukee, WI 53201, USA}
% ...
\affiliation[42]{Bartol Research Institute and Dept. of Physics and Astronomy, University of Delaware, Newark, DE 19716, USA}
% ...
\affiliation[62]{Deutsches Elektronen-Synchrotron DESY, Platanenallee 6, D-15738 Zeuthen, Germany}
% ...

%% file: introduction.tex
\section{Introduction}

In the five decades since Einasto et al.~\cite{Einasto:1974dra} and Ostriker et al.~\cite{Ostriker:1974lna} independently provided evidence that galaxies sat within large halos of unseen matter, the evidence has led to a well-established but troubling conclusion: the luminous matter we observe constitutes only a small fraction of the total matter of the Universe~\cite{deswart:2024xxx}.
This evidence for dark matter (DM) comes from scales on the order of galaxies~\cite{Rubin:1978kmz,Bosma:1981xxx,Mandelbaum:2005nx,Gavazzi:2007vw,Hoekstra:2005kg,Guzik:2002zp}, to galaxy clusters~\cite{Clowe:2006eq,Bardeau:2007ds,Hoekstra:2007nc,Berge:2007mn,Marrone:2009ye,McInnes:2009zv}, and onto that of the Universe~\cite{Planck:2018vyg,2dFGRS:2001csf,SDSS:2005xqv}.
Presently, DM is expected to comprise 84\% of the Universe's total matter content~\cite{Planck:2018vyg}. 
However, despite its seeming ubiquity, DM has eluded a decades-long effort to understand its nature.
Nevertheless, particle DM is well-motivated both experimentally~\cite{Markevitch:2003at} and theoretically~\cite{Bertone:2004pz}, and in this context, weakly interacting massive particles (WIMPs)---particles that interact with normal matter at or below the weak scale---are particularly interesting.
Thus, there is a global, multi-pronged effort to detect such particles: direct searches look for deposited energy from DM scattering in calorimeters~\cite{Liu:2017drf}; production searches look for missing transverse momentum in accelerators~\cite{Buchmueller:2017qhf}; and indirect searches look for the stable byproducts of DM annihilation and decay~\cite{Conrad:2017pms,PerezdelosHeros:2020qyt,Arguelles:2019ouk,Arguelles:2022nbl}.

Since neutrinos can escape the dense astrophysical environments where DM should accumulate, they play an essential role in indirect DM searches.
In fact, searches for neutrinos from DM annihilation have been carried out in the direction of the Sun~\cite{IceCube:2016dgk,IceCube:2025fcu,IceCube:2021xzo,ANTARES:2016xuh,Super-Kamiokande:2015xms,Super-Kamiokande:2011wjy,Super-Kamiokande:2004pou}, the center of the Earth~\cite{IceCube:2024yaw,IceCube:2016aga,ANTARES:2016bxz}, the Galactic Center (GC)~\cite{IceCube:2011kcp,IceCube:2023ies,ANTARES:2019svn,Super-Kamiokande:2020sgt}, dwarf spheroidal galaxies~\cite{Lu:2024jbq}, and nearby galaxy clusters~\cite{IceCube:2013bas}.
The IceCube Neutrino Observatory~\cite{IceCube:2016zyt} is the largest currently operating neutrino observatory, and the data it produces have been used in many such searches.
Within its core, a denser subarray, DeepCore~\cite{IceCube:2011ucd}, is currently deployed lowering the energy threshold down to $\sim$\,\qty{5}{\GeV}. 
However, this energy threshold of the detector has limited the searches to DM with masses $\gtrsim$\,\qty{5}{\GeV}~\cite{IceCube:2021xzo}, so far with peak sensitivity to DM with masses in the range from $\approx$\,\qty{100}{\GeV} to $\approx$\,\qty{1}{\TeV}~\cite{IceCube:2016dgk,IceCube:2025fcu}.

The IceCube Upgrade~\cite{IceCubeCollaborationP:2025rpl} is an infill of the DeepCore subarray, instrumenting the deepest and most transparent glacial ice.
The Upgrade contains more advanced optical modules and enhances IceCube's sensitivity in the \si{\GeV} energy range.
In addition to enhancing IceCube's neutrino oscillation results~\cite{IceCubeCollaborationP:2025rpl}, this new subarray will enable a new generation of DM analyses, opening up new mass regimes and significantly improving sensitivity to lower DM masses.

In this article, we present sensitivity projections for the IceCube Upgrade to neutrino fluxes induced by DM annihilation, focusing on DM masses between \qty{3}{GeV} and \qty{500}{GeV}.
We investigate both solar and GC scenarios and provide comparisons to current limits set by both direct and indirect detection experiments.
Our results indicate that the IceCube Upgrade will significantly improve sensitivities across much of the low-mass DM parameter space within a few years of operation.

The remainder of this article is structured as follows:
\Cref{sec:upgrade} describes the IceCube Upgrade and its key features.
\cref{sec:simulation} details the simulation framework and event selection procedures used in this analysis.
In \cref{sec:analysis_method}, we present the analysis methods, including our statistical approach and the specific treatments for solar and GC DM searches.
The main results and sensitivity projections are given in \cref{sec:results}.
Finally, \cref{sec:conclusion} summarizes our findings and discusses future prospects.

%% file: IceCube_Upgrade.tex
\section{The IceCube Observatory and IceCube Upgrade}
\label{sec:upgrade}

The main array of the IceCube Neutrino Observatory consists of 5{,}160 digital optical modules (DOMs) deployed on 86 strings at depths between \qty{1,450}{\meter} and \qty{2,450}{\meter} within the Antarctic ice.
Each DOM contains one downward-facing photomultiplier tube (PMT) that can detect the Cherenkov radiation produced by charged particles produced in neutrino interactions with atomic nuclei in the ice or underlying bedrock~\cite{IceCube:2016zyt}.
While the main array's primary goal is to measure \si{\TeV} astrophysical neutrinos, the detector also includes DeepCore, a densely instrumented sub-array located deep within the clearest regions of the ice to enable the detection of lower-energy neutrinos.
This enhances IceCube's sensitivity to atmospheric neutrino oscillations~\cite{PhysRevD.108.012014, IceCubeCollaboration:2024ssx}, \si{GeV}-scale physics~\cite{IceCube:2016dgk,IceCube:2021xzo,ArguellesDelgado:2021lek}, low-energy astrophysical signals~\cite{IceCube:2022lnv}, and Earth tomography~\cite{Donini:2018tsg}.
While DeepCore's design specification was to reach an energy threshold of \qty{10}{\GeV}~\cite{IceCube:2011ucd}, recent analyses have been able to lower this threshold to \qty{5}{\GeV}~\cite{IceCubeCollaboration:2024ssx,IceCube:2021xzo}.

\begin{figure}[t]
    \centering
    \includegraphics[width=0.7\linewidth]{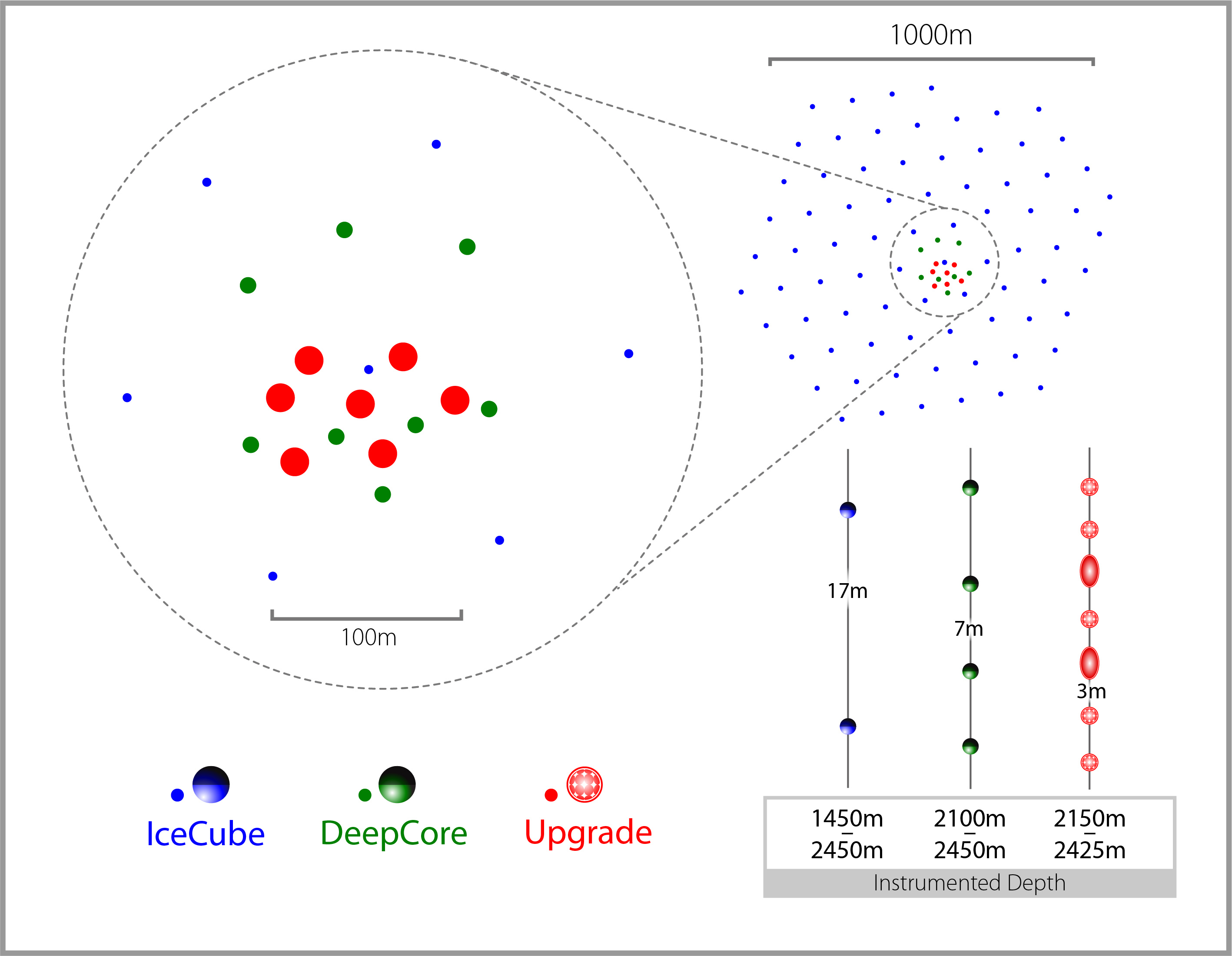}
    \caption{\textbf{\textit{Top-down View of the IC93 detector.}}
    Each point represents one of the strings on which IceCube's optical modules are deployed, with symbols denoting different deployment sub-detectors.
    A magnified view of the more densely instrumented DeepCore and Upgrade region is shown in the top left.
    }
    \label{fig:upgrade_representation}
\end{figure}

The planned IceCube Upgrade, shown schematically in \cref{fig:upgrade_representation}, is a dense, seven-string infill within the DeepCore fiducial volume.
For clarity we will refer to the detector configuration with the Upgrade as IC93 and the previous configuration without the Upgrade as IC86.
The Upgrade strings are instrumented with two new optical modules: the D-Egg~\cite{IceCube:2022mng} and the mDOM~\cite{IceCube:2019anq,IceCube:2021eij,IceCube:2023eij}.
The D-Egg contains two eight-inch PMTs, one pointing vertically upwards, and one downwards, in an oblong pressure housing.
The multi-PMT mDOM contains 24 three-inch PMTs pointing in all directions in a spherical pressure housing.
In total, 280 D-Eggs and 400 mDOMs are distributed evenly across the seven new strings.

In addition to increased per-module photocathode area, these new modules offer more uniform angular coverage and improve the angular resolution in comparison to the previous generation of single-PMT DOMs.
The interstring spacing is reduced from the DeepCore spacing of \qty{40}{\m}--\qty{70}{\m} down to \qty{20}{\m}--\qty{30}{\m}.
Additionally, the new strings are more densely instrumented with optical modules than those in the DeepCore array, with the inter-module vertical spacing reduced from \qty{7}{\m} to just \qty{3}{\m}.
These modules are placed at depths between \qty{2,160}{\m} and \qty{2,430}{\m}, where the ice is most transparent.
The improved modules and denser spacing enhance IceCube's sensitivity to \si{\GeV}-scale neutrinos.


%% file: event_selection.tex
\section{Simulation and Event Selection}
\label{sec:simulation}

\subsection{Detector Simulation}
\label{sec:det_sim}

In this work, we use the simulation and event selection criteria described in Ref.~\cite{IceCubeCollaborationP:2025rpl}; however, we summarize the most essential details of these processes here.

We simulate events from neutrino interactions, atmospheric muons, and noise events resulting from coincident PMT noise in the complete upgraded detector configuration.
Neutrino events are injected using the \texttt{GENIE} software package~\cite{Andreopoulos:2009rq}.
The all-flavor neutrino and anti-neutrino events are sampled from an $E_{\nu}^{-2.5}$ spectrum, to enhance the production of low energy neutrinos, with interaction vertices sampled within a cylinder containing the DeepCore fiducial volume and including both charged-current (CC) and neutral-current (NC) interactions.
The atmospheric muons are simulated with the \texttt{MuonGun} package~\cite{Becherini:2005sr}, which produces muons on a cylinder just outside the IceCube fiducial volume.
The angular and energy distributions of these events are biased towards events that will more likely survive to later selection stages, with the unphysical bias removed via weighting~\cite{LeonardDeHolton:2022tyg}.

Both atmospheric muons and muons produced in muon-neutrino CC interactions are propagated with the \texttt{PROPOSAL} package~\cite{Koehne:2013gpa}, while all other final state particles are propagated with \texttt{GEANT4}~\cite{GEANT4:2002zbu}.
Any photons emitted by these particles are then propagated using the \texttt{CLSim}~\cite{9041727} package using the ice model described in Ref.~\cite{IceCube:2024qxf}.

When a photon strikes the surface of a DOM, it may eject a photoelectron (PE) that initiates a detectable signal. The probability of this occurring depends on the quantum efficiency and, for non-Upgrade modules, the DOM's angular acceptance.
At this point, noise PEs are also added, and PE arriving within \qty{0.2}{\nano\second} are combined into a singular pulse.
The total charge of the PE is sampled from a calibrated charge distribution and pulses with a charge greater than or equal to 0.25 times the nominal PE charge are kept.
The timing information of each pulse is then smeared according to a Gaussian distribution, with an additional global Gaussian smearing applied to all pulses within a module.


There are two distinct triggers for events: DeepCore SMT3~\cite{IceCube:2011ucd,IceCube:2016zyt}, applicable to existing strings, and a newly developed trigger that incorporates the IceCube Upgrade strings.
This new trigger window operates on the concept of a local coincidence (LC), which is achieved when two pulses on the same DOM are recorded in a \qty{10}{\nano\second} window or when two pulses are seen on DOMs within eight modules of each other in a \qty{250}{\nano\second} window.
If eight LCs are seen within a \qty{1.75}{\micro\second} window, the detector triggers and all pulses are read out from \qty{4}{\micro\second} before until \qty{6}{\micro\second} after the trigger condition was met.
Most triggered events are atmospheric muons and pure noise, which trigger the detector at a rate of $\approx$\,\qty{100}{\Hz} and $\approx$\,\qty{10}{\Hz}, respectively. We expect these rates to be reduced with future updates to the trigger.

The first step of the processing is a noise cleaning algorithm that attempts to remove pulses caused by PMTs noise.
The noise removal algorithm is a graph-neural-network-based approach based on the \texttt{DynEdge} model~\cite{Abbasi:2022ypr} and implemented with the \texttt{GraphNeT} package~\cite{Orsoe:2025aqu}.
This approach can cut the number of noise pulses by an order of magnitude while retaining 95\% of signal pulses.
This method dramatically improves over the previous method, described in Ref.~\cite{IceCubeCollaborationP:2025rpl}, by cutting the fraction of noise pulses to 6.8\%, compared to the 70\% achieved with the previous method.

\begin{figure}
     \centering
     \begin{subfigure}[b]{0.48\textwidth}
         \centering
         \includegraphics[width=\textwidth]{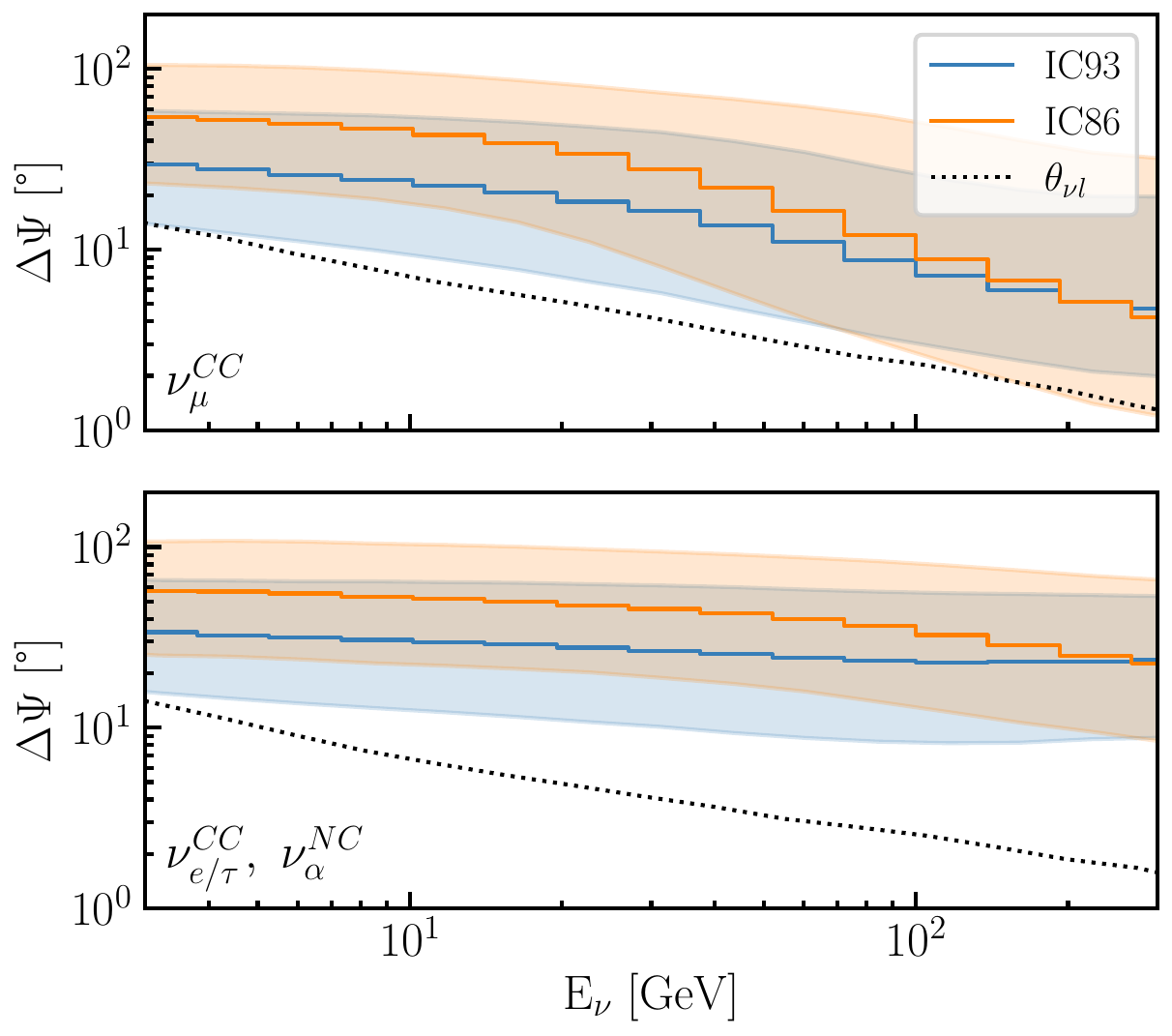}
         
     \end{subfigure}
     \hfill
     \begin{subfigure}[b]{0.48\textwidth}
         \centering
         \includegraphics[width=\textwidth]{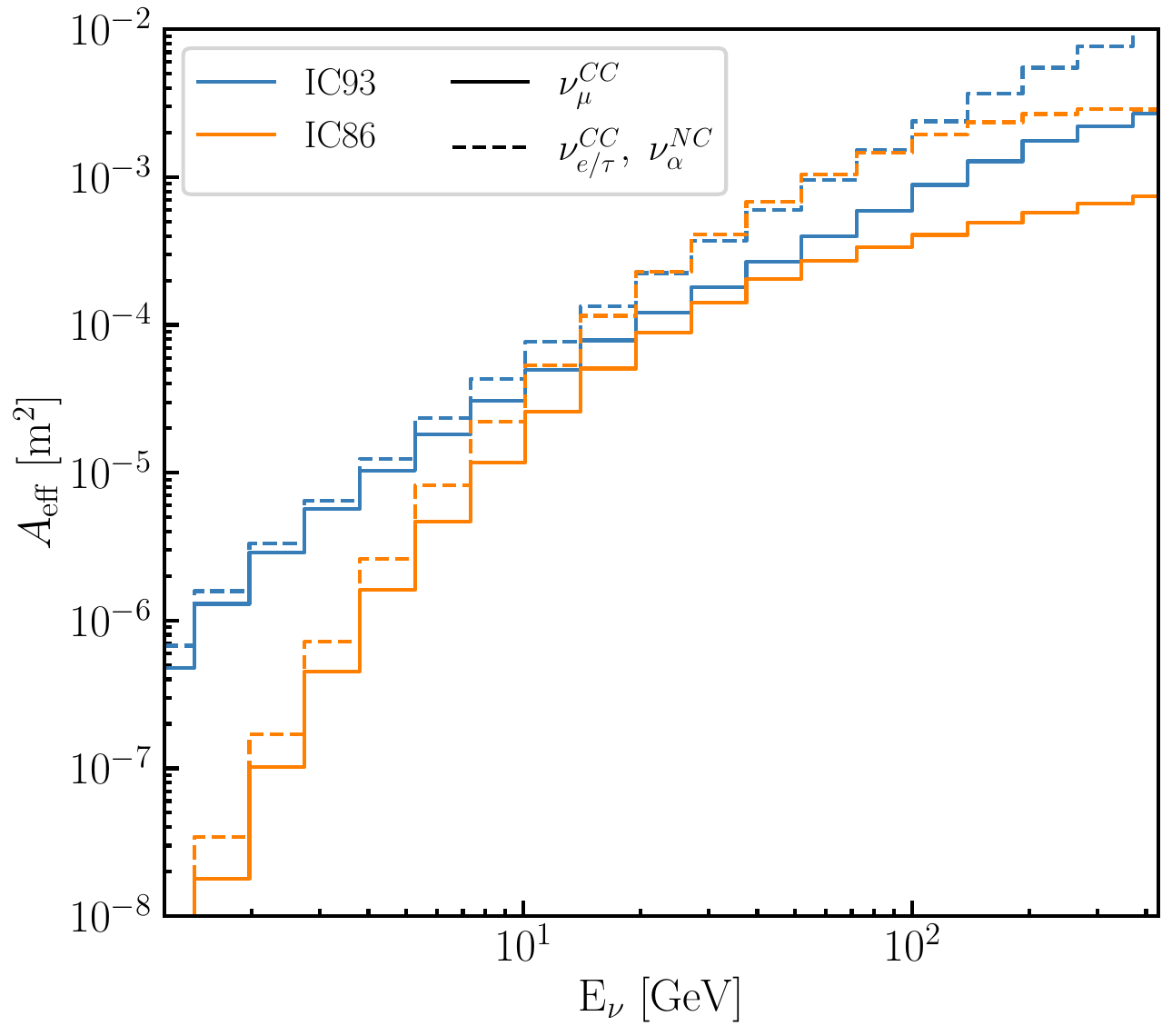}
     \end{subfigure}
        \caption{\textbf{\textit{Detector response comparison between the IC93 and IC86.}}
        The left plots show the angle ($\Delta \Psi$) between the true and reconstructed neutrino direction as a function of neutrino energy.
        The line represents the flavor-averaged median value, while the band shows the $1\sigma$ containment region.
        The black dotted line represents the median opening angle between the produced lepton and the incoming neutrino in CC interactions.
        The right plot shows the flavor-averaged effective area as a function of neutrino energy.
        Together, these yield increased statistics and more precise pointing that enhance the Upgrade's power relative to previous analyses. 
        }
        \label{fig:detector_response}
\end{figure}


The event selection procedure for the upgraded detector begins by addressing the dominance of atmospheric muons and detector noise at the trigger level, which exceeds  neutrino rates by several orders of magnitude.
To proceed in data filtering, multiple reconstruction quantities are defined, but only three are used in the analysis: the reconstructed energy $E_\mathrm{reco}$, reconstructed opening angle to the source $\cos\theta_\mathrm{reco}$, and morphology Track Score. The Track Score ranges from 0 to 1 and provides a quantitative estimate of the event's track-like nature, where higher values indicate a stronger track-like feature.
To obtain a neutrino-enhanced sample, the new upgrade filter, adapted from the DeepCore filter, utilizes a veto region surrounding the DeepCore fiducial volume.
Events are rejected if signals detected in the veto region are temporally and spatially consistent with originating from the same event.
Following this initial filtering, a sequence of five cuts is applied to further suppress background. Four variables are derived from noise-cleaned pulses: the number of DOMs registering pulses, the number of pulses within the fiducial volume, the vertical position ($z$) of the first hit DOM, and the timing difference between the first and last pulses. An additional variable considers the time span of all pulses prior to noise cleaning. These selections, informed by previous DeepCore analyses, reduce pure noise triggers by more than three orders of magnitude, primarily due to the multiplicity-based cuts.
To identify and reject residual atmospheric muon events, the selection incorporates three machine-learning classifiers: a graph neural network and two boosted decision trees utilizing reconstructed event properties and pulse spatial-temporal distributions. The combined application of these classifiers achieves an additional three orders of magnitude reduction in muon background. 
At this stage of the selection, the event sample is neutrino-dominated with no simulated pure noise events surviving the event selection. The neutrino rate is also increased in contrast to previous detector configurations.

These performance improvements can be summarized by two important detector response parameters: the angular resolution and the effective area.
The change in these quantities from the detector configuration without the Upgrade to the configuration with the Upgrade is summarized in \cref{fig:detector_response}. 
The improved angular resolution and the enhancement in event rates, by up to two orders of magnitude at the lowest energies, will not only increase the statistics in the dataset but also significantly reduce background contamination.
These advancements will enhance the sensitivity of searches for neutrinos originating from low-mass DM interactions in astrophysical sources.


\subsection{Dark Matter Simulation}
\label{sec:dm_sim}

Both analyses employ the \charon{} software package~\cite{Liu:2020ckq} to model the neutrino yield from DM annihilation.
In this framework, it is assumed that annihilation proceeds via a single dominant channel, producing two well-defined Standard Model (SM) particles.
In this analysis we consider annihilation to $\bb{}$, $\tautau$, and $\nunu{}$ pairs, which give three broadly representative spectral shapes.
Respectively, they produce a soft neutrino spectrum, a hard neutrino spectrum, and a line-like peak.
The resulting prompt particle spectra are obtained from pre-computed tables.
These SM particles subsequently decay, ultimately yielding neutrinos.
In dense media, such as the solar core, a Monte Carlo (MC) treatment is used to accurately model the interplay of interactions and decay processes that affect the resulting neutrino flux.

\begin{figure}[t]
    \centering
    \includegraphics[width=0.5\linewidth]{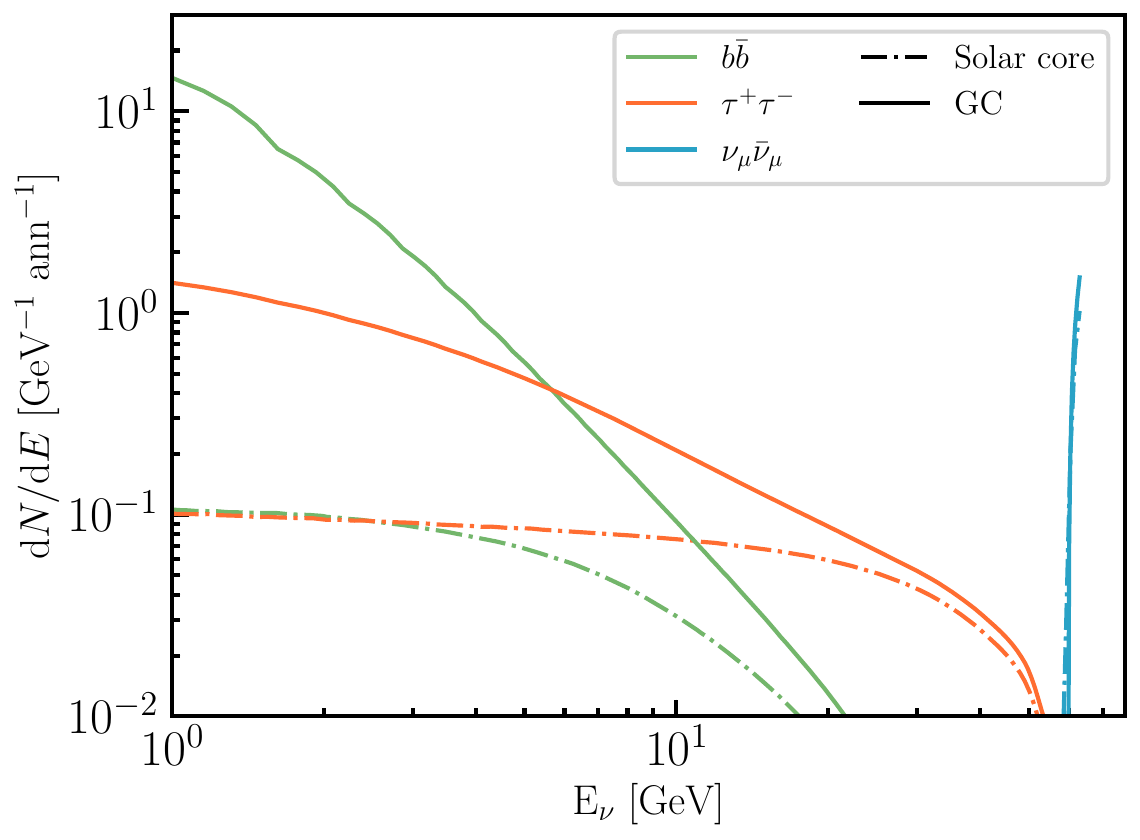}
    \caption{
    \textbf{\textit{Comparison of Signal Fluxes from DM Annihilation in the Solar Core and GC.}}
    Simulated normalized neutrino fluxes
    per annihilation for $m_\chi = \SI{63}{\GeV}$ are shown for three different annihilation channels,
    originating from DM interactions in the GC (solid lines) and solar core (dash-dotted lines).
    }
    \label{fig:signal-production}
\end{figure}

The characteristic shapes of the main signal spectra considered are summarized in \cref{fig:signal-production}.
The differences observed in the low-energy spectra are attributable to the fate of hadrons, particularly charged pions, produced during hadronization.
In the GC, these leptons can decay and contribute additional low-energy neutrinos to the flux, whereas in the solar core, they rapidly lose energy through inelastic scattering leading either to absorption by the mediums nucleis or decaying in neutrinos with energies below the Upgrades energy threshold. 
This therefore suppresses the low-energy neutrino component of the spectrum in the solar core case.

    

%% file: analysis_method.tex
\section{Analysis Methods}
\label{sec:analysis_method}

\subsection{Statistical method}

In both the solar core and GC analyses, the statistical analyses follow a similar workflow, starting from a binned Poisson likelihood to derive the exclusion sensitivity to different annihilation models of DM in the respective sources.

First, signal and background event templates are derived using MC simulations for the IceCube Upgrade detector configuration.
Each template provides the expected event rates in the relevant analysis bins.
For each DM model, characterized by a mass $m_\chi$ and a specific annihilation channel, the signal template is constructed by sampling events from the source direction and weighting by the simulated flux.
The weighted events are then binned according to the analysis-dependent reconstructed quantities, such as reconstructed energy and opening angle to the source.
The background templates are built within the same parameter space, substituting the DM fluxes for expected fluxes from background sources, including atmospheric muons, conventional atmospheric neutrinos, and other significant analysis-dependent backgrounds.

The analyses quantify the data's agreement with particular model parameters using the binned Poisson likelihood, given by:
\begin{equation}
\mathcal{L}(n \mid \boldsymbol{\theta}) = \prod_{i} \frac{\mu^{(i)}(\boldsymbol{\theta})^{n^{(i)}} e^{-\mu^{(i)}(\boldsymbol{\theta})}}{n^{(i)}!},
\label{eq:poisson_likelihood}
\end{equation}
where $n^{(i)}$ is the observed event count in bin $i$, and $\mu^{(i)}(\boldsymbol{\theta})$ is the expected number of events for a given model hypothesis, given by:
\begin{equation}
\mu_i(\boldsymbol{\theta}) = \alpha_{\chi} \, \mu_{\chi}^{(i)} + \alpha_{\mathrm{bg}} \, \mu_{\mathrm{bg}}^{(i)}.
\label{eq:model}
\end{equation}
Here, $\boldsymbol{\theta}=(\alpha_{\chi},\, \alpha_{\mathrm{bg}})$ are the normalizations of the DM and background fluxes with respect to a nominal model and $\mu_{\chi,i}$ and $\mu_{\mathrm{bg},i}$ are the signal and background templates.
Furthermore, the test statistic (TS), given by:
\begin{equation}
\mathrm{TS}(\boldsymbol{\theta_{1}},\,\boldsymbol{\theta_{2}}) = 2\left[ \log \left(\mathcal{L}(n  \mid \alpha_{\chi}^{(1)},\, \alpha_{\mathrm{bg}}^{(1)})\right) - \log \left(\mathcal{L}(n  \mid \alpha_{\chi}^{(2)},\, \alpha_{\mathrm{bg}}^{(2)})\right) \right]\mathrm{,}
\end{equation}
quantifies the data's preference between two models parametrized by $\boldsymbol{\theta}_{1}$ and $\boldsymbol{\theta}_{2}$, namely the background-plus-signal (DM) model and the background-only (no DM) model.
In this case, a positive value indicates a preference for the former, while a negative value indicates a preference for the latter.

We then quantify the Upgrade's performance by calculating the DM model parameters that would be excluded assuming the null hypothesis is true, i.e., assuming no Dark Matter signal ($\alpha_{\chi}=0$).
To do this, pseudodata are generated by Poisson sampling the background-only, i.e. $\alpha_{\chi}=0$, model.
The likelihood is then maximized according to the background-only and background-plus-signal hypotheses.
In practice, bins in which the MC simulation yields no background events, typically located at the extremes of the phase space where detector efficiency is low and the event selection leaves sparse coverage, are excluded from the likelihood evaluation. 
In these bins, the prediction of zero expected events is driven by limited simulation statistics rather than a physical expectation, and the Poisson likelihood is not well-defined for an expectation of zero. These bins are therefore masked in all analyses
Best fit values for these models---given by $\hat{\boldsymbol{\theta}}=(\hat{\alpha}_{\chi},\, \hat{\alpha}_{\mathrm{bg}})$ and $\check{\boldsymbol{\theta}}=(0,\, \check{\alpha}_{\mathrm{bg}})$ ---thus give a test statistic $\mathrm{TS}(\hat{\boldsymbol{\theta}},\, \check{\boldsymbol{\theta}})$ which gives the preference for the background-plus-signal model over the background-only model.
Since the latter model lives in a strict subset of the former, this quantity is non-negative.


The background test-statistic distribution is then derived from repeating this procedure many times.
Wilks' theorem~\cite{Wilks:1938dza} states that with a large number of samples this test-statistic distribution should follow a $\chi^2$ distribution with one degree of freedom.
In these analyses, however, the distribution follows a modified chi-squared distribution~\cite{Cowan:2010js} with one degree of freedom.
This modification arises from the requirement that the model normalizations are non-negative.
Whereas in the standard $\chi^2$ for 1 degree of freedom 90\% of the sampled tests statistics are below a value of 2.71, this critical value changes to 1.64 in the truncated distribution.
This critical value has been verified in both analyses using $\approx\!10^4$ pseudoexperiments.

Having computed this critical value, the exclusion normalization, $\alpha_{\chi}^{\mathrm{excl.}}$, for a particular set of pseudodata is then defined by solving:
\begin{equation}
    1.64 = \mathrm{TS}(\hat{\boldsymbol{\theta}},\, (\alpha_{\chi}^{\mathrm{excl.}}, \hat{\alpha}_{\mathrm{bg}}))\mathrm{,}
\end{equation}
where $\hat{\boldsymbol{\theta}}$ is the best-fit point for the background-plus-signal model.
This procedure is then repeated many times, and the median of the resulting distribution is the analysis exclusion sensitivity.




\subsection{Solar WIMP Annihilation}

This analysis assumes that captured WIMPs in the solar core reach capture-annihilation equilibrium, in which the annihilation rate is determined solely by the capture rate.
For sufficiently light WIMPs, evaporation from the Sun can disrupt this equilibrium~\cite{Gould:1987ju}.
The evaporation mass, below which this effect becomes significant over the lifetime of the Sun, is approximately $3.7\,\mathrm{GeV}$~\cite{Gould:1987ju}.
We adopt this as the lower bound on the WIMP mass in the solar analysis for all channels, unless a higher bound is imposed by the mass of the annihilation products.
Using the fitting functions of~\cite{Busoni:2013kaa}, we verify that at $m_\chi = 3.7\,\mathrm{GeV}$ and the cross sections probed in this analysis, the ratio of the evaporation rate to the annihilation rate is below $10^{-3}$, confirming that evaporation is negligible at this threshold.

For this analysis, MC events are binned in three dimensions to construct sampled distributions for both the signal neutrinos originating from DM annihilation in the solar core and the background, which includes atmospheric muons as well as conventional and solar atmospheric neutrinos.
The variables used for binning are: the reconstructed energy of the interacting neutrino, the reconstructed angular separation from the Sun, and the track score.
The latter serves as a morphological classifier designed to distinguish \(\nu_{\mu}\) charged-current interactions from other event topologies and takes values in [0, 1] with 1 describing the events most similar to tracks.
The events are binned in 36 linear bins running from $0^\circ$ to $180^\circ$ in the opening angle, $12$ logarithmic bins running from 3 to \qty{300}{GeV} in energy and $3$ bins in track score from 0 to 1.

\begin{figure}[t]
    \centering
    \includegraphics[width=1.0\linewidth]{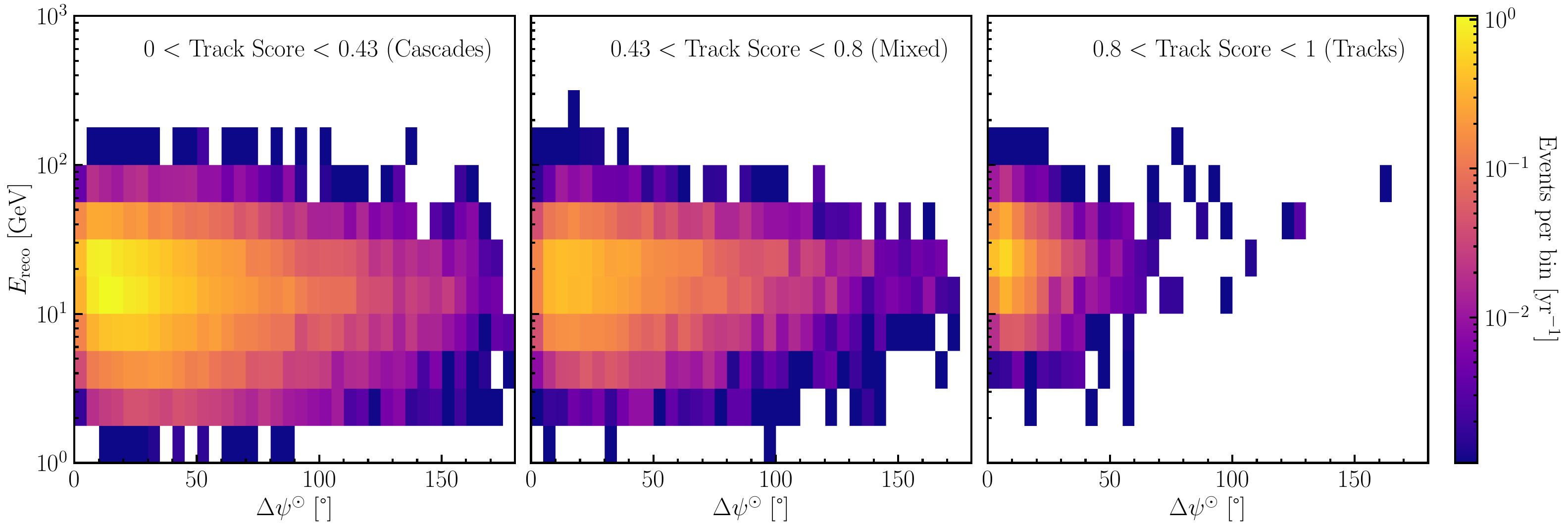}
    \caption{\textbf{\textit{Example signal distribution for solar DM.}}
    The distribution is simulated for DM with mass $m_{\chi} = 63 \ \mathrm{GeV}$ and scattering cross section $\sigma_{\chi p} = \qty{e-40}{\cm\squared}$ annihilating to $\bb{}$, from the core of the sun.
    Event rates distribution binned in reconstructed energy, opening angle ($\Delta\psi^{\odot}$) to the Sun, and reconstructed track score.}
    \label{fig:signal_pdf_m1_sun}
\end{figure}

To construct the signal distributions, event times are randomly sampled within a three-year window.
For each sampled time, the position of the Sun is calculated using an internal IceCube software tool that interfaces with the Positional Astronomy Library~\cite{Jenness2013}, taking the Modified Julian Date and the geographic coordinates of the detector as inputs.
Events are retained only if their true direction falls within the solar disk, which has an angular extent of approximately ~0.5° as seen from Earth~\cite{refId0}.
Each retained event contributes a rate proportional to the product of its MC weight and the DM flux, which is computed using the \charon{} tool, to construct the desired distribution~\cite{Liu:2020ckq}. 

This selection process is applied at multiple randomly selected epochs across the three-year window, with the results accumulated to generate a smoothed physical distribution of neutrino events originating from DM annihilation in the core of the Sun accurately taking into account the variation in solar positions across a calendar year.
The distribution is constructed using a reference cross-section of \(\sigma_{\chi p} = \qty{e-40}{\cm\squared}\).
An example of this distribution is shown in Fig. \ref{fig:signal_pdf_m1_sun}.
As expected, the distribution exhibits a significant spread in reconstructed direction, reflecting the challenges associated with angular reconstruction at these energies.
Furthermore, events with higher track scores, e.g., \(\nu_{\mu}\) events, display a narrower spread as expected from track morphologies.

The construction of the background distribution varies depending on the component being considered.
For the solar atmospheric background component, which consists of neutrinos produced when cosmic rays interact in the solar atmosphere, the same time-dependent selection methodology applied to the signal flux is used~\cite{Arguelles:2017eao}.
In contrast, a different approach is used for the conventional atmospheric and muon background distributions. 

\begin{figure}[t]
    \centering
    \includegraphics[width=1.0\linewidth]{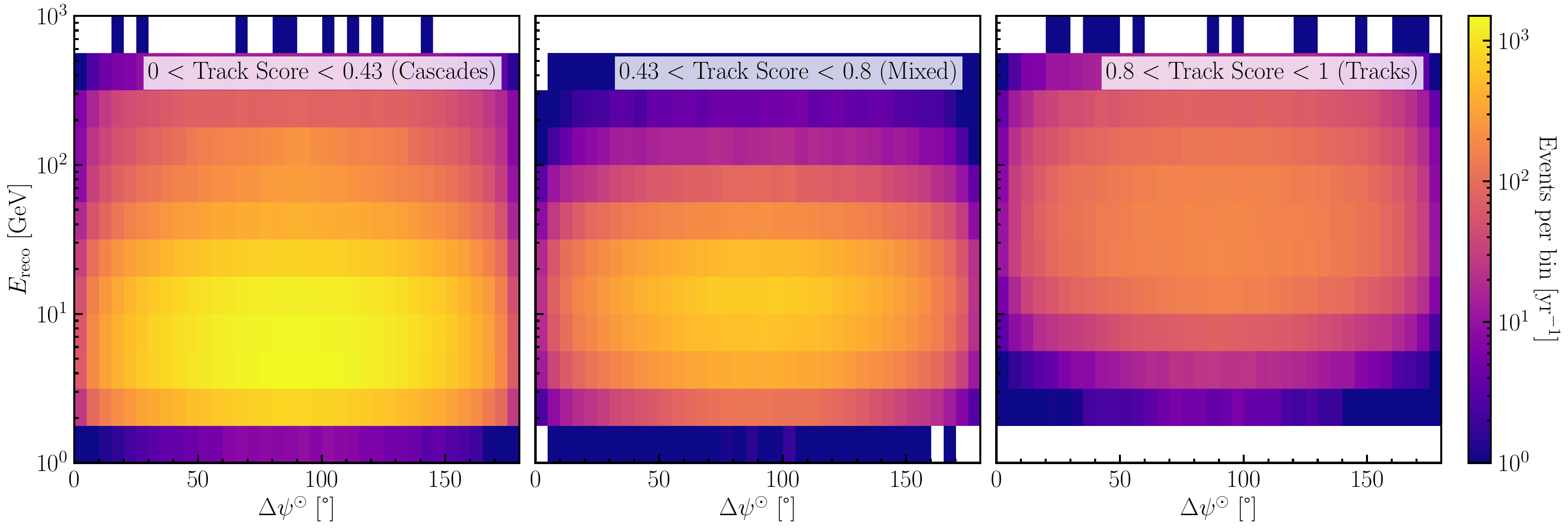}
    \caption{\textbf{\textit{Background distribution for solar analysis.}}
    The simulation includes events arising from solar and conventional atmospheric neutrinos and atmospheric muons.
    All events from detector noise are removed at the final analysis level.
    }
    \label{fig:bgk_pdf_sun}
\end{figure}

The background distribution is typically constructed by reshuffling the right ascension of the detector data, where random right ascension values are drawn uniformly from the range \([0, 2\pi)\), and the events are binned in the relevant variables.
This procedure is repeated multiple times to ensure a sufficient sample size for binning, with the resulting distributions averaged to form the final background distribution.
However, since this is a prospective study and no detector data from the IceCube Upgrade configuration were available, we rely on the MC simulations of the atmospheric neutrino and muon background.
All neutrino MC events are weighted to the flux from Ref.~\cite{Honda:2015fha}, and muon events are weighted as described in \cref{sec:det_sim}.
A simulated detector dataset is then generated by sampling events according to Poisson statistics based on these rates, with the right ascension randomized by drawing values uniformly from the interval \([0, 2\pi)\).
The sum of these three components forms the total background distribution, which is shown in \cref{fig:bgk_pdf_sun}.


For this analysis, we evaluated detector systematic uncertainties on the results. 
The variations include shifts in the optical efficiency of the standard IceCube and DeepCore DOMs, where the DOM efficiency was increased or decreased by 10\% relative to the nominal value, as well as analogous variations applied to the Upgrade DOMs, for which the efficiency was shifted by $\pm 5\%$. 
In addition to DOM‑related effects, we tested modifications to the bulk ice model by varying the scattering and absorption lengths by $\pm 10\%$ relative to baseline.
Running the full analysis with these systematically varied samples shows that the resulting impact on the sensitivity remains at the level of $\sim$\,20\% or less compared to the nominal MC, well below the expected $1\sigma$ statistical fluctuation. 
Consequently, the results presented in this work rely on the nominal simulation.

\subsection{Galactic Center}
In this work, the GC analysis adopts the binned Poisson likelihood in two observables: reconstructed energy and opening angle to the GC while both cascade-like and track-like topologies are included.
The approach is in analogy with the previous work for IceCube DeepCore in~\cite{IceCube:2025fcn}.

The neutrino flux resulting from DM annihilation in the GC is computed as a function of neutrino energy $E_\nu$ and the opening angle from the GC $\psi$ as follows:

\begin{figure}[t]
    \centering
    \includegraphics[width=1.\linewidth]{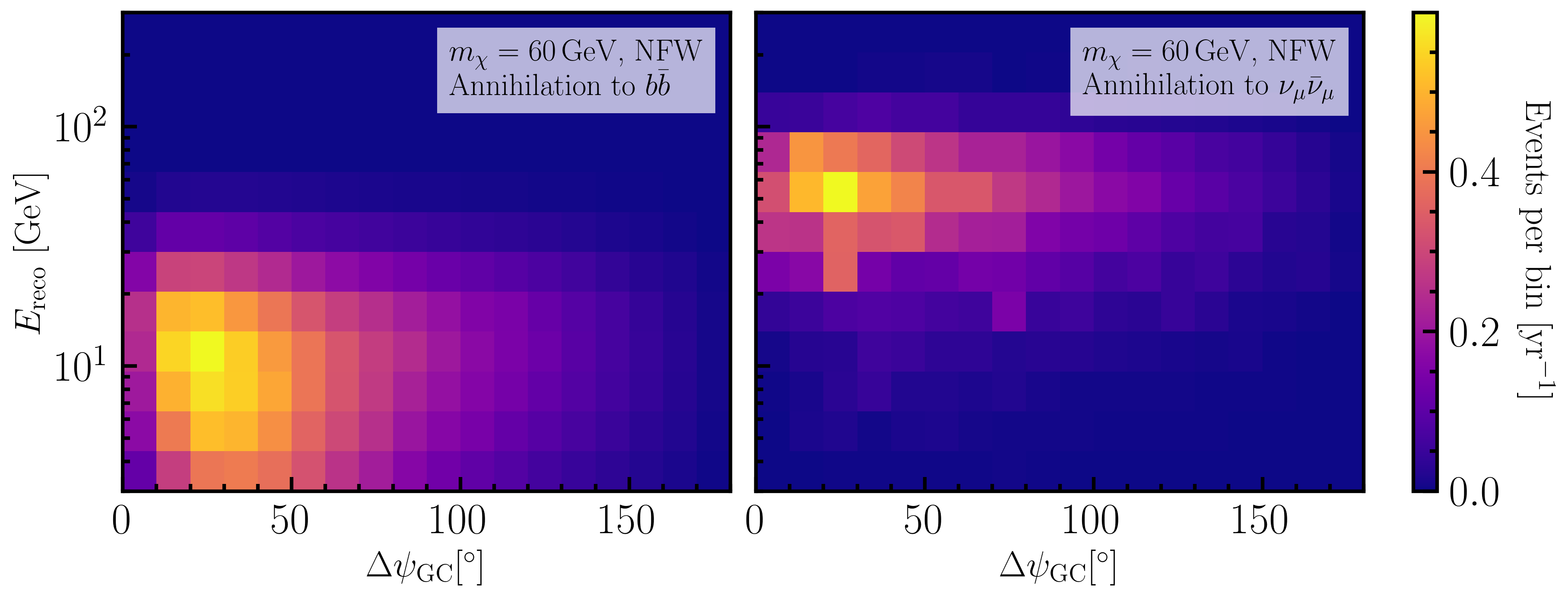}
    \caption{\textbf{\textit{Example signal distribution for GC DM.}}
    The distribution is simulated for DM with mass $m_{\chi} = 60 \ \mathrm{GeV}$ and thermally averaged cross-section $\langle \sigma \upsilon \rangle=\qty{e-24}{\cm\cubed\per\second}$ annihilating to $\bb{}$ (left) and $\nu_\mu \bar{\nu}_\mu$ (right), assuming the NFW halo profile.
    Event rates distribution binned in reconstructed energy and opening angle from the GC $\Delta \Psi_{GC}$.
    }
    \label{fig:signal_pdf_m1_gc}
\end{figure}

\begin{align}
     \frac{\mathrm{d^2}\phi_{\nu}}{\mathrm{d}E_{\nu} \mathrm{d} \Omega} (E_\nu, \psi) &=  \frac{1}{4\pi} \, \frac{\langle \sigma \upsilon \rangle}{\mathit{k} \, m_{\mathrm{DM}}^2} \; \frac{\mathrm{d}N_{\nu}}{\mathrm{d}E_{\nu}} \int_{\mathrm{LoS}} \rho^2 \left(r(l, \psi) \right) \; \mathrm{d} l \; , 
   \label{eq:GCsignal_flux}
\end{align}

where $\langle \sigma \upsilon \rangle$ is the thermally averaged annihilation cross-section. $\Omega$ is the solid angle that can be expressed in terms of the opening angle $\psi$.
The factor $k$ depends on the nature of the DM particles: $k=4$ for Dirac particles and $k=2$ for Majorana ones.
In this work, we assume that DM is a Majorana particle~\cite{Queiroz:2016sxf}, corresponding to the more optimistic signal flux.

The term $\frac{\mathrm{d}N_\nu}{\mathrm{d}E_\nu}$ is the differential number of neutrinos produced per DM annihilation per neutrino energy and is computed using \charon{}.
This spectrum is propagated to the Earth, assuming an averaged oscillation over a very long baseline, with the oscillation parameters taken from NuFit 5.2~\cite{Esteban:2020cvm}.

The rightmost integral of \cref{eq:GCsignal_flux}, commonly referred to as the \textit{J-factor}, encapsulates the astrophysical contribution from the DM halo. It is taken along the line-of-sight (LoS) and over the solid angle $\Omega$. 
 The density distribution of the DM as a function of the radius from the GC is $\rho(r)$, which is assumed to be spherically symmetric.
 
The precise morphology of the Galactic DM halo has been a subject of ongoing debate for decades, as the DM density in the central region of the Milky Way remains observationally unconstrained~\cite{Benito:2019ngh}.
Cosmological N-body simulations of collision-less cold DM tend to predict a \textit{cusp} profile for DM halos, with density rising sharply toward the center of galaxies~\cite{Navarro:1995iw}. In contrast, observational evidence from rotation curves of dwarf galaxies suggests a \textit{core} profile, where the density remains approximately constant~\cite{Burkert_1995}. Modern DM halo simulations that include baryonic feedback suggest that the large gravitational potential of baryonic matter in the cores of Milky Way–like galaxies can further steepen the DM density into a pronounced cuspy profile \cite{Gnedin:2004cx, Gustafsson:2006gr}. On the other hand, in smaller galaxies, feedback processes associated with star formation and supernovae are expected to dominate, diluting and hence flattening the inner density, leading to the formation of a roughly constant-density core \cite{DiCintio:2013qxa}.
In this projection study, we adopt a common benchmark choice which is a parametric representative for a cusp profile referred to as the Navarro-Frenk-White (NFW) profile~\cite{Navarro:1995iw}.
The corresponding parameter values of the Milky Way are extracted based on the motion data within the galactic region as in~\cite{Nesti:2013uwa}.
The J-factor integration is then computed using the {\sc Clumpy} package~\cite{Hutten:2018aix}.

The binned signal expectation is then constructed by convolving the signal flux \eqref{eq:GCsignal_flux} with a \textit{response matrix} built from MC simulation.
This response matrix characterizes how neutrinos, defined by their true type (i.e., neutrino or antineutrino and flavor), energy, and incident angle, are detected and reconstructed by the experimental apparatus.
We also apply a kernel density estimation (KDE)~\cite{KDEpy} to smooth the matrix and reduce the fluctuation caused by limited MC statistics.

\begin{figure}[t]
    \centering
    \includegraphics[width=0.7\linewidth]{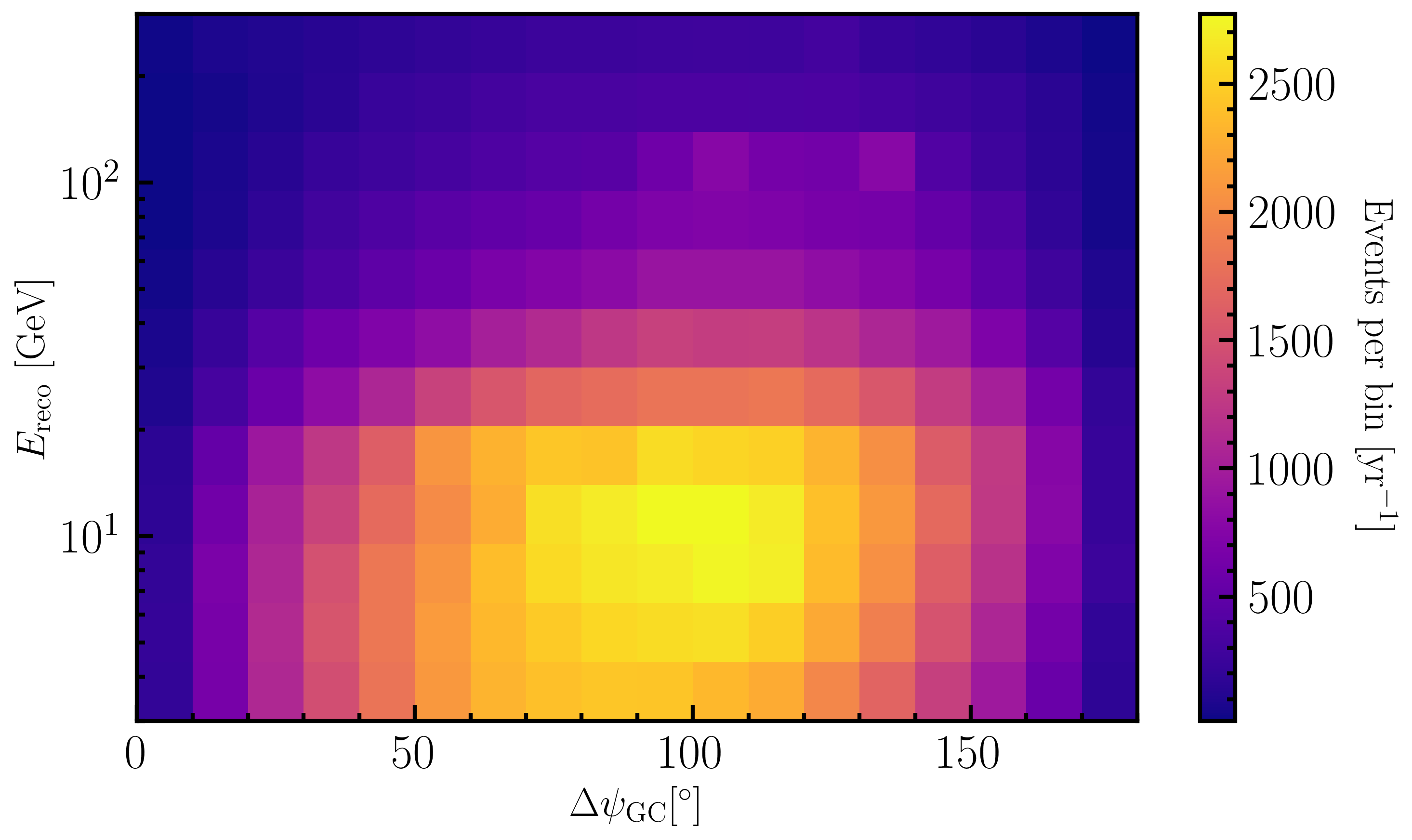}
    \caption{\textbf{\textit{Background distribution for GC analysis.}}
    The simulation includes events arising from conventional atmospheric neutrinos and atmospheric muons.
    All events from detector noise are removed at the final analysis level.
    }
    \label{fig:bgk_pdf_gc}
\end{figure}

\cref{fig:signal_pdf_m1_gc} presents two examples of the expected signal, shown as two-dimensional distributions of events with respect to reconstructed energy and opening angle from the GC.
These are made assuming annihilation of DM with $m_\chi$ = \qty{60}{GeV} into a pair of $\bb{}$ and $\nu_\mu \bar{\nu}_\mu$ in the case of NFW halo profile.
Both distributions exhibit a pronounced excess in the region close to the GC.
The events are binned in 18 linear bins running from $0^\circ$ to $180^\circ$ in the opening angle and $12$ logarithmic bins running from 3 to \qty{300}{GeV} in energy.
The energy distribution of the $\bb{}$ channel features an excess that spreads to the energies below the DM mass value due to its soft power-law energy spectra.
On the other hand, the $\nu_\mu \bar{\nu}_\mu$ channel exhibits a peak centered on the DM mass, due to the line-like neutrino spectrum discussed in \cref{sec:dm_sim}.

In the energy range considered for this analysis---\qty{3}{\GeV}--\qty{300}{\GeV}---the background is dominated by atmospheric muons and neutrinos.
The atmospheric background expectation is constructed by re-weighting the MC events following the atmospheric neutrino flux taken from~\cite{Honda:2015fha}.
The same KDE technique used for the signal computation is also applied to calculate the background distribution, which results in the two-dimensional background event distribution shown in \cref{fig:bgk_pdf_gc}.

Another possible background is the diffuse neutrino flux caused by the interaction of cosmic rays with the medium of the galactic disk.
The flux has been measured by IceCube at TeV energies and above~\cite{IceCube:2023ame}. At lower energies, in the GeV regime, this component has not yet been measured, as existing searches typically focus on energies above $\approx$\,1~TeV \cite{IceCube:2017trr, IceCube:2019lzm, IceCube:2023ame, ANTARES:2022izu}, where current detectors have sufficient sensitivity. Therefore, an estimation has been made based on both models and measurements from Ref.~\cite{IceCube:2023ame}.
The energy spectra are extrapolated down to the GeV region assuming an unbroken power law spectra.
This reference derives the spatial distribution of the neutrino flux from the diffuse gamma-ray flux measured by Fermi~\cite{Fermi-LAT:2012edv}.
To do this, they assume that both originate from the same interactions between cosmic rays and the interstellar medium.
This estimation results in a negligible contribution, on the order of 0.01\%-0.05\%, to the total number of expected events in the energy range of the analysis.
Different galactic plane neutrino emission models \cite{Gaggero:2015xza} show a $\sim$\,10\% variation in the sensitivities, well below the $1 \sigma$ statistical uncertainties.
For that reason, we do not consider this background component in the final results presented \cref{sec:results}.

This analysis examines the impact of detector systematics in the same manner as the solar analysis.
As in that case, detector systematics are found to have a subdominant impact on the sensitivity relative to statistical fluctuations.

The primary source of uncertainty for this analysis comes from the DM profiles.
It can result in an order of magnitude change in the sensitivities between the cusp (NFW) and cored (Burkert) profiles~\cite{RevModPhys.93.035007,IceCube:2023ies}.
This work considers the NFW profile, representing the most optimistic scenario.

%% file: results.tex
\begin{figure}[t]
    \centering
    \includegraphics[width=0.7\linewidth]{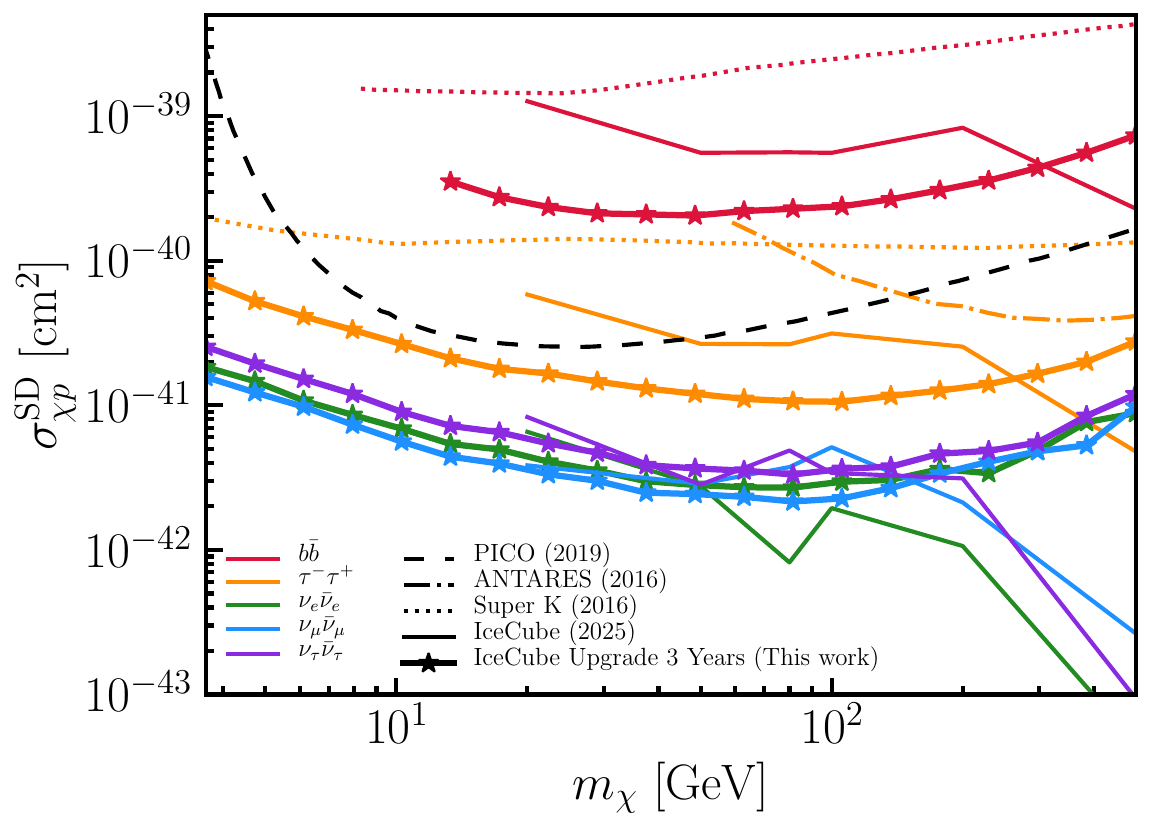}
    \caption{\textbf{\textit{Sensitivity to Solar DM annihilation: DM-proton scattering $\sigma_{\chi p }$ with three years of data.}} These projected limits of the IceCube Upgrade extend IceCube's limits to $3.7\,\mathrm{GeV}$, offering complementarity to the lowest masses probed by direct detection experiments. Furthermore, the IceCube Upgrade is expected to produce leading limits on DM-proton scattering cross sections below $\mathcal{O}(\SI{100}{\GeV})$. The prospective limits from the IceCube Upgrade are compared to limits from Ref.~\cite{KM3Net:2016zxf,Super-Kamiokande:2015xms,PICO:2019vsc}.}
    \label{fig:Solar_DM_sensitivity}
\end{figure}

\section{Results}
\label{sec:results}

For the solar core study,
the projected sensitivity of the IceCube Upgrade to DM annihilation signals is summarized in \cref{fig:Solar_DM_sensitivity} and compared with the most recent results of similar studies from other leading DM experiments~\cite{KM3Net:2016zxf,Super-Kamiokande:2015xms,PICO:2019vsc,IceCube:2025fcu}. Here the IceCube Upgrade is projected to become the most sensitive experiment to low-mass DM annihilation for masses up to approximately \SI{200}{\GeV} for the $\bb{}$ and $\tautau{}$ channels, and up to approximately \qty{20}{\GeV} for neutrino final states.
The lowest WIMP mass tested for each annihilation channel is constrained by two different factors.
In the case of $\nunu{}$ and $\tautau{}$ channels, the limiting factor is the assumption of negligible WIMP evaporation from the solar core, which is only valid for WIMP masses above $\sim\!3.7\,\mathrm{GeV}$~\cite{Gould:1987ju}.
For the $\bb{}$ channel, the limitation arises from the use of \texttt{PYTHIA}, which does not reliably simulate hadronic decays for $\bb{}$ systems with invariant masses below $\sim$\,\qty{10}{\GeV} due to the breakdown of the approximations it relies on, such as the assumption of a continuum of final states and the use of the string-fragmentation model, in the hadronic resonance region; see Ref.~\cite{Sjostrand:2014zea}.

\begin{figure}[t]
    \centering
    \includegraphics[width=0.7\linewidth]{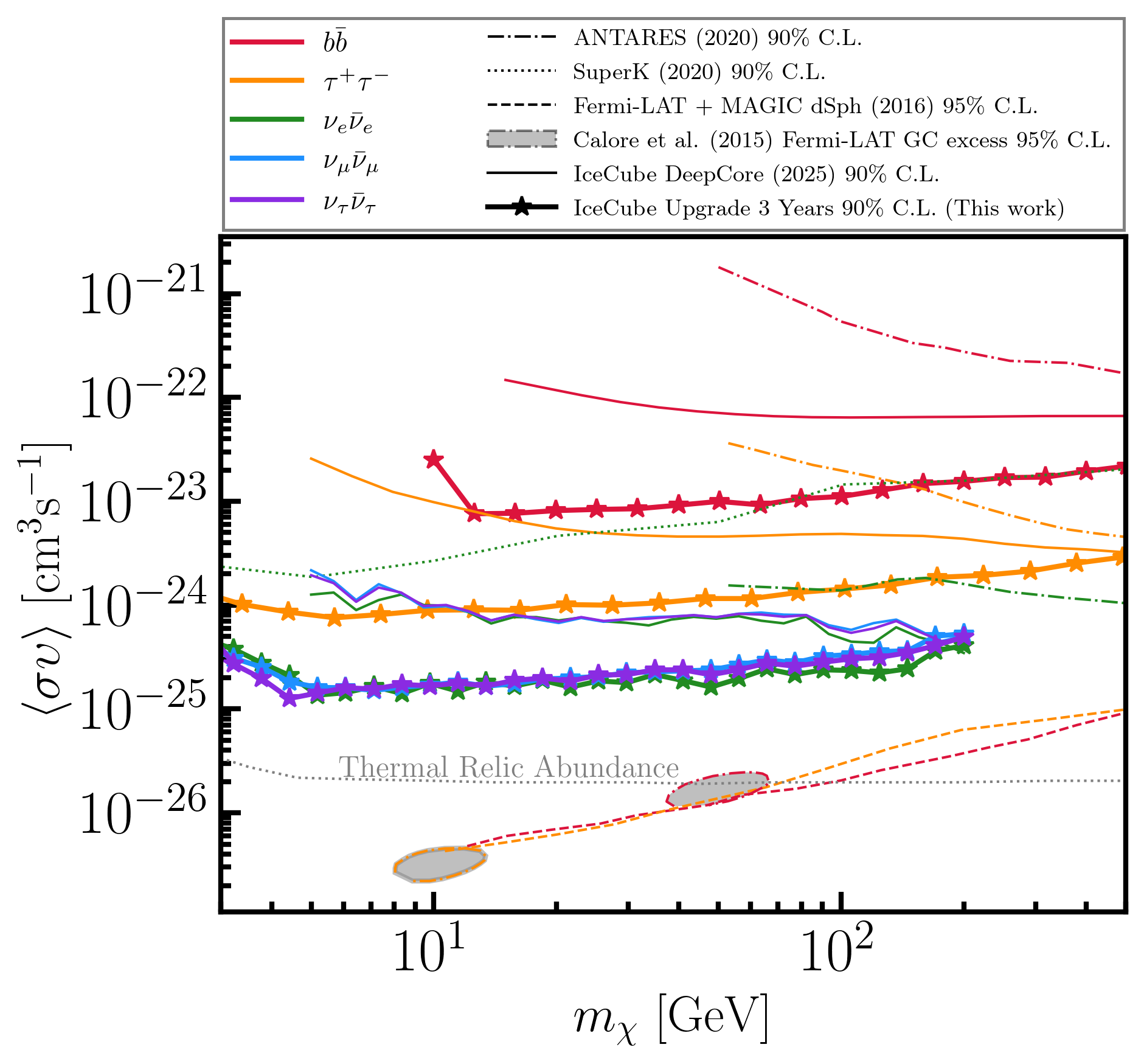}
    \caption{\textbf{\textit{Sensitivity to Galactic Center DM annihilation: thermally averaged cross section with three years of data.}}
    The projected sensitivity shows that the IceCube Upgrade is expected to improve over previous IceCube limits across the entire mass range considered in this work.
    Furthermore, the IceCube Upgrade will significantly improve limits on annihilation to neutrino pairs.
    The prospective limits from the IceCube Upgrade are compared to limits from Refs.~\cite{IceCube:2023ies,ANTARES:2019svn,Super-Kamiokande:2020sgt,MAGIC:2016xys,Calore:2014nla}. The cross-section required to produce the observed relic abundance from thermal freeze-out is taken from Ref.~\cite{Steigman:2012nb}.
    }
    \label{fig:gc_sensitivity}
\end{figure}

The projected sensitivity of the IceCube Upgrade to DM annihilation signals from the GC is presented in \cref{fig:gc_sensitivity}, assuming the NFW halo profile. We also compare the sensitivities obtained in this work with results of other experiments~\cite{IceCube:2023ies,ANTARES:2019svn,Super-Kamiokande:2020sgt,MAGIC:2016xys,Calore:2014nla}. As illustrated, Upgrade is expected to match the current limits of the 9.3-year DeepCore dataset within just three years of operation. Notably, for DM masses below \qty{20}{\GeV}, the IceCube Upgrade achieves an order of magnitude improvement.
In the case of neutrino line channels, the projected sensitivities reach within a factor of a few of the thermal relic annihilation cross section on the order of $\sim\!\SI{e-25}{\cm^3 \s^{-1}}$, establishing competitive constraints in the mass range between \qty{3}{\GeV} and \qty{500}{\GeV}.

%% file: conclusion.tex
\section{Conclusion}
\label{sec:conclusion}
We have presented a comprehensive sensitivity study for DM annihilation searches using the IceCube Upgrade, targeting both the solar core and the GC as primary sources.
Our analysis employs a binned Poisson likelihood framework, leveraging MC simulations to construct signal and background templates across multiple observables, including reconstructed energy, angular separation, and track score. 

For the GC, our projections demonstrate that the IceCube Upgrade will achieve sensitivities to the thermally averaged DM annihilation cross section comparable to, and in several mass ranges exceeding, the most stringent existing limits with three years of data. 
Since this analysis is not background free, the sensitivity evolves proportionally to $t^\frac{1}{2}$, where $t$ is the livetime.
Notably, for DM masses below 20~GeV and for neutrino line final states, the Upgrade is expected to improve exclusion limits by up to an order of magnitude, establishing leading sensitivity in this mass window.

In the solar analysis, the IceCube Upgrade will substantially extend the experiment's sensitivity to lower DM masses, achieving constraints on the DM-proton scattering cross section down to $3.7\,\mathrm{GeV}$.
These projected limits offer complementarity to the lowest masses currently probed by direct detection experiments.
For annihilation to both hadronic and leptonic final states, the expected sensitivities improve upon the most stringent existing results, with the Upgrade anticipated to set leading limits on DM-proton scattering cross sections below $\approx\!100~\mathrm{GeV}$.

Our study demonstrates that the IceCube Upgrade will provide significant advances in the search for DM-induced neutrino signals. These results emphasize the transformative potential of the IceCube Upgrade for indirect DM detection, motivating continued development of both theoretical modeling and event reconstruction in the next generation of neutrino telescopes.

%% file: acknowledgments.tex
\acknowledgments
The authors gratefully acknowledge the support from the following agencies and institutions:
USA {\textendash} U.S. National Science Foundation-Office of Polar Programs,
U.S. National Science Foundation-Physics Division,
U.S. National Science Foundation-EPSCoR,
U.S. National Science Foundation-Office of Advanced Cyberinfrastructure,
Wisconsin Alumni Research Foundation,
Center for High Throughput Computing (CHTC) at the University of Wisconsin{\textendash}Madison,
Open Science Grid (OSG),
Partnership to Advance Throughput Computing (PATh),
Advanced Cyberinfrastructure Coordination Ecosystem: Services {\&} Support (ACCESS),
Frontera and Ranch computing project at the Texas Advanced Computing Center,
U.S. Department of Energy-National Energy Research Scientific Computing Center,
Particle astrophysics research computing center at the University of Maryland,
Michigan State University,
Astroparticle physics computational facility at Marquette University,
NVIDIA Corporation,
and Google Cloud Platform;
Belgium {\textendash} Funds for Scientific Research (FRS-FNRS and FWO),
FWO Odysseus and Big Science programmes,
and Belgian Federal Science Policy Office (Belspo);
Germany {\textendash} Bundesministerium f{\"u}r Forschung, Technologie und Raumfahrt (BMFTR),
Deutsche Forschungsgemeinschaft (DFG),
Helmholtz Alliance for Astroparticle Physics (HAP),
Initiative and Networking Fund of the Helmholtz Association,
Deutsches Elektronen Synchrotron (DESY),
and High Performance Computing cluster of the RWTH Aachen;
Sweden {\textendash} Swedish Research Council,
Swedish Polar Research Secretariat,
Swedish National Infrastructure for Computing (SNIC),
and Knut and Alice Wallenberg Foundation;
European Union {\textendash} EGI Advanced Computing for research;
Australia {\textendash} Australian Research Council;
Canada {\textendash} Natural Sciences and Engineering Research Council of Canada,
Calcul Qu{\'e}bec, Compute Ontario, Canada Foundation for Innovation, WestGrid, and Digital Research Alliance of Canada;
Denmark {\textendash} Villum Fonden, Carlsberg Foundation, and European Commission;
New Zealand {\textendash} Marsden Fund;
Japan {\textendash} Japan Society for Promotion of Science (JSPS), Ministry of Education, Culture, Sports, Science and Technology (MEXT), and Institute for Global Prominent Research (IGPR) of Chiba University;
Korea {\textendash} National Research Foundation of Korea (NRF);
Switzerland {\textendash} Swiss National Science Foundation (SNSF).

%% file: appendix.tex
\appendix

%% file: references.tex
\bibliographystyle{unsrt}  
\bibliography{references}